\documentclass[%
 aip,
 amsmath,amssymb,
 reprint,%
]{revtex4-1}

\usepackage{graphicx}% Include figure files
\usepackage{dcolumn}% Align table columns on decimal point
\usepackage{bm}% bold math
\usepackage[utf8]{inputenc}
\usepackage[T1]{fontenc}
\usepackage{mathptmx}
\usepackage{etoolbox}

\makeatletter
\def\@email#1#2{%
 \endgroup
 \patchcmd{\titleblock@produce}
  {\frontmatter@RRAPformat}
  {\frontmatter@RRAPformat{\produce@RRAP{*#1\href{mailto:#2}{#2}}}\frontmatter@RRAPformat}
  {}{}
}%
\makeatother
\begin{document}

\preprint{AIP/123-QED}

\title{A cold-insertable scanning probe microscope for dry dilution refrigerators with picometer stability and ultra-low electron temperatures}
% Force line breaks with \\
\author{Y. Wei}
 \affiliation{Department of Physics, The Hong Kong University of Science and Technology, Clear Water Bay, Kowloon, Hong Kong SAR}%Lines break automatically or can be forced with \\
\author{B. J\"ack}%
 \email{bjaeck@ust.hk}
\affiliation{Department of Physics, The Hong Kong University of Science and Technology, Clear Water Bay, Kowloon, Hong Kong SAR}%

\date{\today}% It is always \today, today,

\begin{abstract}
Investigating the microscopic mechanisms of quantum materials requires high-resolution scanning probe techniques, often based on atomic force microscopy (AFM). However, implementing AFM in cryogen-free dilution refrigerators with picometer stability is challenged by intense pulse tube mechanical noise ($S_{\rm PT}(f)\approx10^{-6}\,\text{m}/\sqrt{\text{Hz}}$) and the fundamental trade-off with thermalizing piezoelectric motion stages below 100~mK. Here, we demonstrate an AFM-based scanning microwave impedance microscope integrated onto a standard cold-insertable probe that successfully resolves this bottleneck. By employing a two-pronged design strategy---coupling a mechanically stiff AFM module with a magnetic-field-compatible, critically damped internal spring-suspension---we achieve an extremely low relative tip-sample vibration noise density of $S_{\rm AFM}(f)<10^{-11}\,\text{m}/\sqrt{\text{Hz}}$. This yields a spectrally integrated relative tip-sample displacement of $\Delta z\approx10\,\text{pm}$, representing a greater than 100-fold stability improvement over recent fast-loading dry SPM setups. Simultaneously, optimized thermal interfaces, customized copper strapping, and comprehensive RF filtering enable a local sample electron temperature of $T_{\rm e}\leq 60\,$mK, circumventing the thermal penalties of mechanical decoupling. By avoiding permanent structural modifications to the host cryostat, this robust, modular architecture provides an accessible framework for adapting other scanning probe techniques, accelerating the exploration of fragile quantum phases in dry cryostats with picometer stability.
\end{abstract}

%\begin{abstract}
%Investigating the microscopic mechanisms of quantum materials requires high-resolution scanning probe techniques, often based on atomic force microscopy (AFM). However, implementing AFM in cryogen-free dilution refrigerators with picometer stability is challenged by intense pulse tube mechanical spectral noise density $S_{\rm PT}(f)\approx10^{-6}\,\text{m}/\sqrt{\text{Hz}}$ and the difficulty of thermalizing piezoelectric motion stages below 100~mK. Here, we demonstrate an AFM-based scanning microwave impedance microscope in a dry refrigerator achieving an extremely low relative tip-sample motion with a spectral noise density of $S_{\rm AFM}(f)<10^{-11}\,\text{m}/\sqrt{\text{Hz}}$. By combining a stiff, home-built AFM module with a critically damped spring-isolation stage, we realize a spectrally integrated displacement of $\Delta z\approx10\,\text{pm}$. Furthermore, optimized thermal interfaces, thermal strapping, and comprehensive RF filtering yield a low electron temperature of approx. $60\,$mK, despite the mechanical decoupling of the AFM from the mixing chamber. Relying on fundamental design principles rather than complex commercial components or modifications to the cryostat structure, this modular approach can be readily adapted to other techniques, such as scanning SQUID-on-tip or nitrogen-vacancy center microscopy, advancing the study of quantum materials in cryogen-free environments.
%\end{abstract}

\maketitle

\section{\label{sec:intro}Introduction}

\begin{figure}[htbp]
    \centering
    \includegraphics[width=8.5cm]{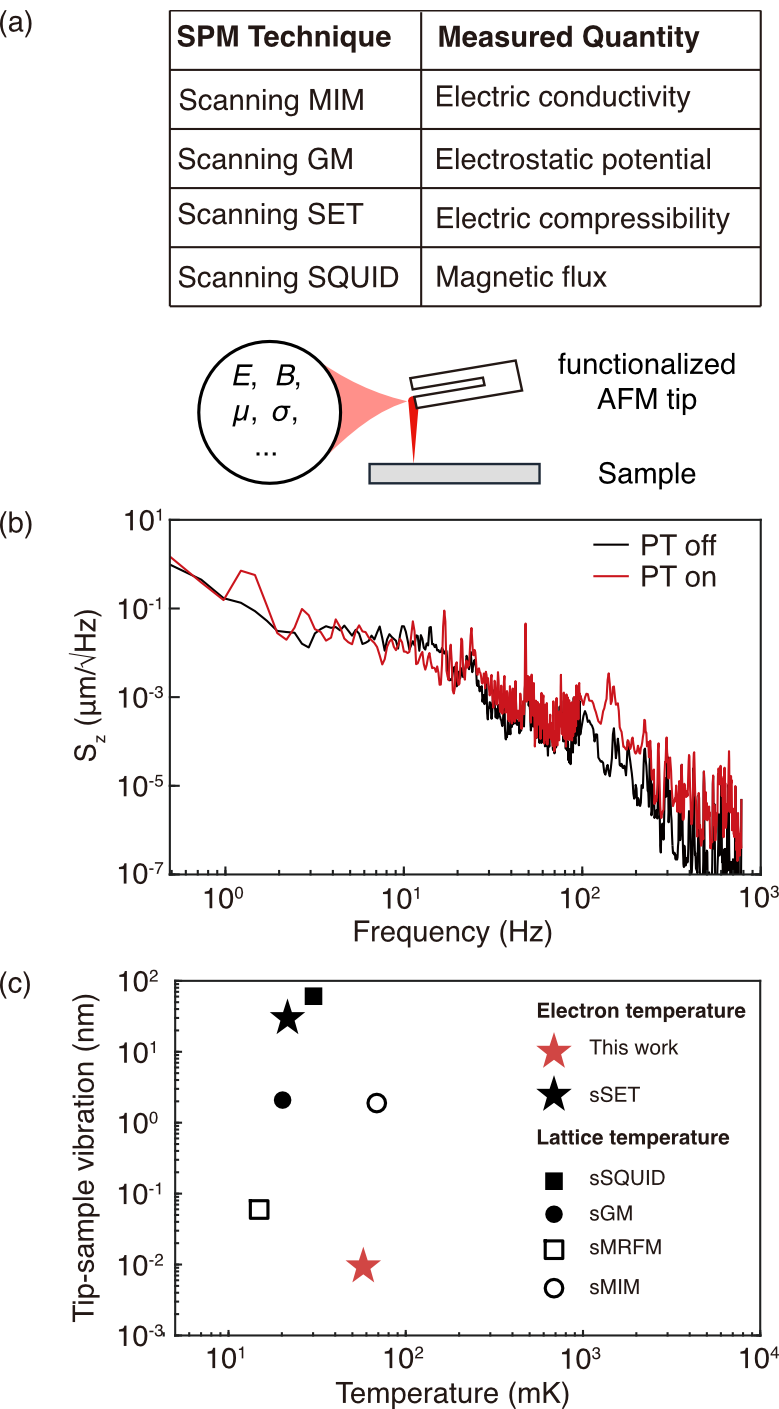}
    \caption{\textbf{Overview of representative AFM-based scanning probe techniques and their vibration, and temperature performance.} \textbf{(a)} Representative atomic force microscopy (AFM)-based scanning probe techniques for locally measuring diverse physical properties of novel materials, including electric fields \(E\), magnetic fields \(B\), chemical potential \(\mu\), electrical conductivity \(\sigma\), and other material properties. \textbf{(b)} Frequency-dependent vertical displacement noise spectral density measured on the mixing-chamber plate of a Leiden Cryogenics CS-SF 110 dilution refrigerator with the pulse tube cooler (PT) switched on (red) and off (black). \textbf{(c)} Overview of the temperatures and relative tip-sample vibration amplitudes reported for representative dilution refrigerator based scanning probe microscopy techniques [sSET~\cite{barber2024characterization}, sSQUID~\cite{low2021scanning}, sSGM~\cite{pelliccione2013design}, scanning magnetic force microscopy (sMRFM)~\cite{den2014atomic}, sMIM~\cite{cao2023millikelvin}]. Reported electron temperatures are shown as star symbols, reported lattice temperatures are shown by other symbols.}
    \label{fig:fig1}
\end{figure}

The macroscopic properties of quantum materials are inextricably linked to their microscopic electronic, electric, and magnetic characteristics~\cite{keimer2017physics}. Elucidating the underlying mechanisms of these phenomena demands experimental techniques capable of measuring physical observables with high spatial resolution. To this end, scanning probe microscopy (SPM) techniques---particularly those based on the atomic force microscope (AFM)---have emerged as indispensable tools. Techniques such as scanning microwave impedance microscopy (sMIM)~\cite{barber2022microwave}, scanning superconducting quantum interference device (sSQUID) microscopy~\cite{persky2022studying}, scanning gate microscopy (sGM)~\cite{pelliccione2013design, oh2021cryogen}, or scanning single electron transistor microscopy (sSET)~\cite{Barber2024} rely on precise AFM feedback to map quantities such as local conductivity and magnetic fields at the nanoscale~\cite{Telford2023} [see Fig.~\ref{fig:fig1}(a)].

Because many quantum phases---such as fractional quantum Hall states, unconventional superconductivity, and fragile magnetic orders---are characterized by small energy scales, it is often necessary to operate these scanning probe techniques at cryogenic temperatures below one kelvin. Historically, ultra-low temperature experiments were conducted in conventional ``wet'' dilution refrigerator cryostats based on liquid helium dewars. However, the rising costs and periodic interruptions associated with liquid helium consumption have driven a transition toward cryogen-free (``dry'') dilution refrigerators~\cite{Barber2024}. While dry systems offer continuous, low-maintenance operation, they introduce severe technical challenges for realizing sensitive scanning probe techniques~\cite{pelliccione2013design, denHaan2014, Ge2025}.

The most formidable hurdle in adapting AFM to dry dilution refrigerators is the intense mechanical vibration generated by the valve cycle of the Pulse Tube (PT) cryocooler~\cite{Lee2022}. The resulting mechanical noise spectral density measured at the mixing chamber plate of a dilution refrigerator is typically of the order $S_{\rm PT}(f)\approx10^{-6}\,\text{m}/\sqrt{\text{Hz}}$ [see Fig.~\ref{fig:fig1}(b)], which is several orders of magnitude higher than the sub-Angstrom stability required for high-resolution AFM. The spectral noise density is dominated by the fundamental frequency of the mechanical valve cycle, typically on the order of 1-2\,Hz, but also disperses to higher frequencies above 100\,Hz severely affecting SPM measurements~\cite{barber2024characterization}. To date, achieving sub-Angstrom stability in dry systems has historically required extensive structural modifications to the cryostat itself---such as mechanically decoupling the PT coolers~\cite{pelliccione2013,den2014atomic}, complex active/passive damping networks~\cite{pelliccione2013, den2014atomic, geaney2019near}, or highly rigid, thermally insulating architectures~\cite{Ge2025}. However, such approaches preclude the use of convenient, fast-loading sample exchange mechanisms and often compromise the thermal linkage between the mixing chamber and the microscope. Consequently, efficiently thermalizing piezoelectric motion stages and wiring becomes exceptionally difficult, making it challenging to cool the sample and the local environment to electron temperatures below 100~mK [only reported by one dry dilution refrigerator based SPM study to date~\cite{barber2024characterization}] while maintaining low relative tip-sample vibrations [Fig.~\ref{fig:fig1}(c)].

In this article, we demonstrate an AFM-based sMIM in a dry dilution refrigerator that successfully overcomes these mechanical and thermal bottlenecks without necessitating complex structural modifications to the host cryostat. By pairing a mechanically stiff, home-built AFM module (designed to spectrally separate structural eigenmodes from the PT noise) with a critically damped, magnetic-field-compatible spring-isolation stage mounted directly on a standard cold-insertable probe, we suppress the relative tip-sample vibration noise density to $S_{\rm AFM}(f)<10^{-11}\,\text{m}/\sqrt{\text{Hz}}$, yielding a spectrally integrated displacement of $\Delta z\approx10\,\text{pm}$. Furthermore, optimized thermal interfaces, robust thermal strapping, and comprehensive DC filtering achieve a local electron temperature of $T_{\rm eff}\leq60\,$mK despite the mechanical decoupling between the AFM module and the cryostat. This robust, modular architecture readily adapts to other scanning probe techniques, such as SQUID-on-tip or nitrogen-vacancy microscopy, significantly expanding the experimental horizons for studying quantum materials using fast-turnaround, cryogen-free environments.

\section{\label{sec:design}Design of the AFM Module and Vibration Isolation Scheme}

To realize picometer-stable AFM in a dry dilution refrigerator [schematically shown in Fig.~\ref{fig:fig2}(a)], we must simultaneously address two distinct mechanical noise sources. The first consists of building-borne and impact vibrations, characterized by a displacement-noise spectral density $S_{\rm B}(f)$ that dominates the $10^0-10^2\,$Hz range. In conventional ``wet'' cryostats, these external vibrations are the primary concern and are effectively mitigated by isolating the cryostat frame using active or passive room-temperature damping stages~\cite{assig201310}. 

However, cryogen-free systems introduce a second, severe internal noise source: the pulse tube (PT) cryocooler. The PT noise spectral density, $S_{\rm PT}(f)$, is dominated by the fundamental valve-cycle resonance between 1 and 2\,Hz, accompanied by a dispersive broadband spectrum extending from 100 to 1000\,Hz [see Fig.~\ref{fig:fig1}(b)]. Unmitigated, this drives an integrated vertical vibration amplitude of roughly $1\,\mu$m at the mixing chamber (MXC) plate, leading to substantial relative tip-sample vibrations exceeding $10^0\,$nm in typical SPM setups~\cite{barber2024characterization}. Consequently, achieving sub-Angstrom stability requires a dedicated internal noise mitigation scheme that isolates against the broadband hundred-hertz mechanical noise while remaining resilient against excitation by the massive 1--2\,Hz PT fundamental resonance.

Here, we present the design and integration of a tuning-fork-based AFM mounted on a cold-insertable probe inside a \textit{Leiden Cryogenics CF-CS110} dry dilution refrigerator (8\,mK base temperature, equipped with two 1.5\,W PT cryocoolers~\cite{PTcooler} and a 9\,T single-axis solenoid magnet). Our approach to mitigating internal noise relies on a rigorous two-pronged strategy: (1) engineering a stiff AFM module to spectrally separate its mechanical resonance modes from the PT noise spectrum, and (2) implementing a critically damped, magnetic-field-compatible spring-isolation stage at the MXC level. Together, these design choices enable picometer AFM stability at an effective electron temperature of approximately 60\,mK, while fully preserving the convenience of the motorized top-loading mechanism for fast sample exchange.

\subsection{AFM Module with high Mechanical Stiffness} 

Implementing the first prong of our strategy requires pushing the structural eigenmodes of the microscopy module well above the dispersive PT noise spectrum. Most existing low-temperature scanning probe setups~\cite{pelliccione2013design, finkler2012scanning, cao2023millikelvin, Barber2024} rely on commercially available piezoelectric coarse-motion and flexure-based scanners. Because these commercial stages typically possess mechanical resonances below 1\,kHz, they are highly susceptible to acoustic PT noise, resulting in substantial relative tip-sample vibrations often exceeding 30\,nm~\cite{Barber2024}. 

\begin{figure*}[htbp]
    \centering
    \includegraphics[width=17cm]{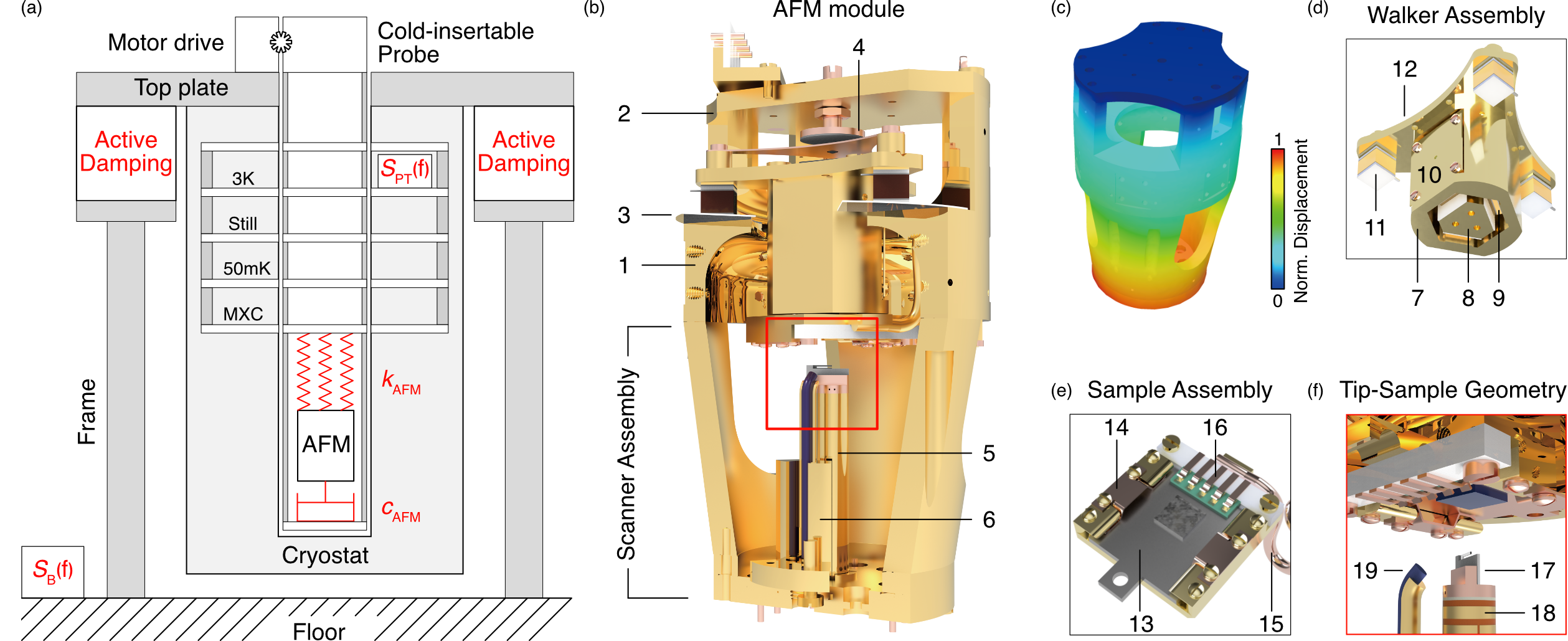}
    \caption{\textbf{Assembly and vibration-isolation architecture of the ultralow-noise, millikelvin AFM integrated into a cryogen-free dilution refrigerator.} \textbf{(a)} Schematic of the complete instrument and its multistage vibration-isolation architecture. The dominant vibration sources are building-borne vibrations, characterized by the displacement-noise spectral density \(S_{\mathrm{B}}(f)\), and vibrations generated by the pulse-tube (PT) cryocoolers, characterized by \(S_{\mathrm{PT}}(f)\). Both contributions are jointly attenuated by a combination of active vibration-isolation stages and an internal passive isolation system at the mixing-chamber (MXC) stage. In the passive system, the suspension springs provide an effective spring constant \(k_{\mathrm{AFM}}\), whereas a home-built damping mechanism provides the damping coefficient \(c_{\mathrm{AFM}}\); together, they form a critically damped spring-mass isolation stage for the AFM module. \textbf{(b)} Cross-sectional CAD rendering of the AFM module and its detachable scanner assembly. The labeled components are the OFHC Cu module body (1), top plate (2), polished \(\mathrm{Al_2O_3}\) planes for \(XY\)-coarse motion (3), spring-force adjustment screw (4), piezoelectric tube scanner~\cite{Ptube} (5), and microwave impedance-matching network (6). \textbf{(c)} Finite-element-method calculation of the normalized-displacement mode shape of the lowest structural eigenmode of the AFM module, corresponding to a horizontal bending mode at \(f_{\mathrm{AFM}} \approx 2.2~\mathrm{kHz}\). \textbf{(d)} CAD rendering of the three-axis \(XYZ\) coarse-positioning walker assembly, comprising the outer \(Z\)-walker body (7), the OFHC Cu prism with sapphire plates (8), shear piezoelectric actuators~\cite{Zpiezo} (9), a CuBe leaf spring (10), three two-axis shear piezoelectric actuators~\cite{XYpiezo} (11), and \(XY\)-walker body (12). \textbf{(e)} Sample assembly comprising a flag-style sample plate (13), two Cu leaf springs (14) providing mechanical clamping and thermal contact, a flexible OFHC Cu braid (15), and five Cu leaf-spring contacts (16). \textbf{(f)} Magnified view of the tip-sample geometry, showing the tuning-fork AFM unit (17), the ring electrode (18) used to excite the tuning fork, and the electrochemically etched PtIr tip (19) connected to the impedance-matching network [(6) in panel (b)].}
    \label{fig:fig2}
\end{figure*}

To achieve sub-nanometer resolution, we bypass commercial stages and employ an extended, highly rigid Pan-style walker architecture~\cite{pan1999he, white2011stiff, battisti2018definition} combined with a modular, high-resonance scanner assembly. A cross-sectional cut of the computer aided design (CAD) of the AFM module is shown in Fig.~\ref{fig:fig2}(b) and the component details (1-19) are described in the figure caption. Furthermore, the design strictly utilizes materials with high thermal conductivity to ensure millikelvin compatibility.

{\em Microscopy Module Body:} The microscopy module is engineered to maximize structural eigenmodes. While sapphire offers an ideal combination of stiffness ($E\approx400\,\text{GPa}$) and thermal conductivity for compact SPM setups~\cite{white2011stiff}, the necessarily larger footprint of our three-axis horizontal coarse-motion module (diameter 48\,mm, height 75\,mm) made sapphire impractical. Consequently, we machined all structural body parts from oxygen-free highly conductive copper (OFHC Cu), which provides excellent thermal conductivity ($>100\,\text{W}/\text{m}\cdot\text{K}$ at 4\,K~\cite{valois2024characterization}). To compensate for copper's comparatively lower Young's modulus ($E\approx120\,\text{GPa}$), we utilized finite element method (FEM) simulations to optimize the module's mechanical geometry. By adopting a conical body structure~\cite{ast2008design, assig201310} with thick walls and compact dimensions [see Fig.~\ref{fig:fig2}(b)], the lowest structural eigenmode---a horizontal bending mode---is pushed to $f_{\rm AFM}\approx2.2\,\text{kHz}$ [Fig.~\ref{fig:fig2}(c)]. This safely separates the module's resonance from the PT noise spectrum.

{\em Coarse Motion and Sample Assembly:} To position the AFM tip over millimeter distances, the sample stage employs an extended three-axis ($XYZ$) Pan-walker architecture [Fig.~\ref{fig:fig2}(d)] integrated directly into the module body. The vertical $Z$-motion and horizontal $XY$-motion are actuated by shear piezoelectric stacks clamped against polished ($R_a<1\,$nm) sapphire plates via CuBe leaf springs. This configuration offers a robust vertical travel range of 5\,mm and horizontal navigation within a 2.5\,mm radius. 

The sample assembly itself [Fig.~\ref{fig:fig2}(e)] is designed to meet three simultaneous criteria: rapid sample exchange, excellent thermalization, and the provision of multiple electrical contacts for device physics (e.g., transport measurements on Hall bars or field-effect transistors). We achieved this using a detachable, flag-style sample plate. Upon insertion, the plate is mechanically clamped by copper leaf springs, ensuring robust thermal anchoring to the OFHC Cu module body via a flexible copper braid, while separate leaf springs establish electrical contact with the sample PCB.

{\em Scanner and Tip Assembly:} To facilitate the fabrication and exchange of complex tuning-fork-tip assemblies---such as those required for sMIM~\cite{cui2016quartz} or scanning SQUID-on-tip~\cite{finkler2012scanning}---the scanner and tip assembly [Fig.~\ref{fig:fig2}(b)] is designed to be easily detachable from the module body. We utilized a long, thin PZT-5A piezoelectric scan tube (length 27.5\,mm, diameter 5\,mm, thickness 0.5\,mm)~\cite{Ptube} as an optimal compromise. It provides a sufficiently large scan range ($\Delta x\approx6\,\text{µm}$ at $T=4\,$K) while maintaining its lowest mechanical eigenmode (a bending mode) at $f_{\rm scanner}\approx1.8\,$kHz, safely above our 1\,kHz threshold. 

The tuning-fork assembly mounts to the tube scanner via a custom banana-plug mechanism [Fig.~\ref{fig:fig2}(f)]. It features a commercial tuning fork ($f_{TF}=32.768\,$kHz) with an electrochemically etched PtIr tip (10\,µm diameter) attached to the free prong, wired to a coaxial cable for sMIM functionality~\cite{giessibl2019qplus, cui2016quartz}. Notably, even with the attached tip wire, the tuning fork consistently exhibits high quality factors of $Q\approx5000-10000$ at cryogenic temperatures ($\leq 3\,$K)~\cite{jiang2023implementing}.

\begin{figure}[htbp]
    \centering
    \includegraphics[width=8.5cm]{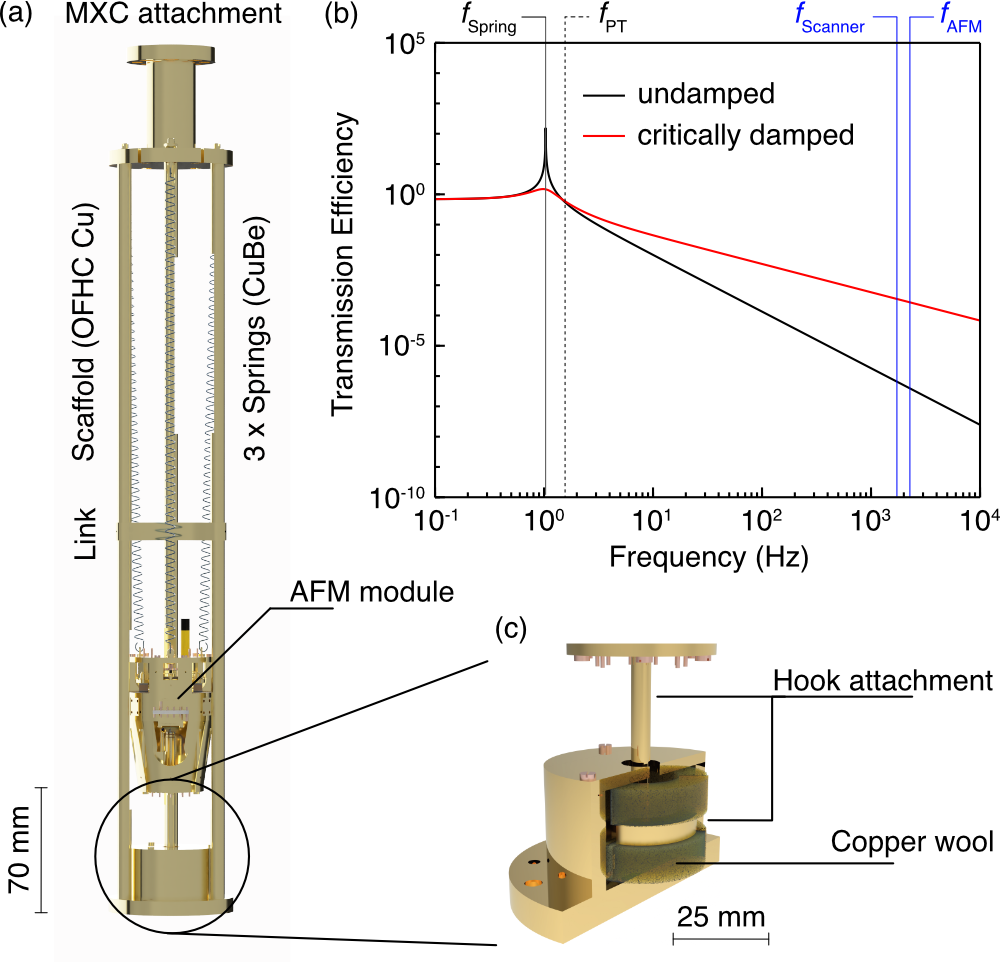}
    \caption{\textbf{Design of a mixing chamber attachment for vibration-isolation design of the AFM module.} \textbf{(a)} Shown is the MXC attachment comprising the AFM module, which is suspended by three CuBe springs providing the effective spring constant \(k_{\mathrm{AFM}}\) and coupled to a home-built damper providing the damping coefficient \(c_{\mathrm{AFM}}\), together forming a critically damped isolation stage. The MXC attachment is realized using a scaffold to support the damping stage at the bottom and comprises a link structure for enhanced mechanical rigidity and as a thermalization point. \textbf{(b)} Calculated frequency-dependent vibration transmission efficiency. The characteristic frequencies are the suspension resonance \(f_{\mathrm{Spring}}\approx1~\mathrm{Hz}\), pulse-tube frequency \(f_{\mathrm{PT}}\approx1.7~\mathrm{Hz}\), the lowest structural resonance of the piezotube scanner \(f_{\mathrm{Scanner}}\approx2.2~\mathrm{kHz}\) and AFM module \(f_{\mathrm{AFM}}\approx2.2~\mathrm{kHz}\), respectively.}
    \label{fig:fig3}
\end{figure}

\subsection{Magnetic-Field-Compatible Vibration Isolation Scheme} 

Having engineered the AFM module with structural resonances well above $1\,$kHz ($f_{\rm scanner}\approx1.8\,$kHz and $f_{\rm AFM}\approx2.2\,$kHz), the second prong of our strategy is to mechanically isolate this module from the remaining environmental and PT noise. 

External building vibrations and impact sound ($S_{\rm B}$) are successfully suppressed using a pair of active vibration isolation stages~\cite{VibIso} installed between the cryostat base plate and the laboratory frame [see Fig.~\ref{fig:fig2}(a)], providing an isolation efficiency of $\approx98\%$ above $10\,$Hz. However, isolating the AFM module from the internal PT noise ($S_{\rm PT}$), which can exceed $100\,\text{nm}/\sqrt{\text{Hz}}$ at the MXC level [Fig.~\ref{fig:fig1}(b)], is substantially more challenging. 

To achieve this, we designed an MXC attachment module [Fig.~\ref{fig:fig3}(a)] that suspends the AFM module (mass $0.45\,$kg) using three custom-made $250\,$mm long CuBe springs (wire thickness $0.6\,$mm), yielding a low suspension resonance of $f_{\rm Spring}\approx1\,$Hz. This soft suspension provides an excellent isolation efficiency of $>99.9\%$ above $1000\,$Hz [Fig.~\ref{fig:fig3}(b)], effectively protecting the high-frequency structural eigenmodes of the AFM from the broadband PT noise. However, because this spring resonance ($1\,$Hz) is spectrally close to the fundamental frequency of the PT valve cycle ($f_{\rm PT}=1.7\,$Hz), the spring is highly susceptible to resonant excitation. Previous attempts using low-frequency springs or flexures~\cite{pelliccione2013design, denHaan2014, geaney2019near, oh2021cryogen, cao2023millikelvin} often suffered from under-damping, resulting in severe relative tip-sample vibrations of $>1\text{--}10\,$nm. 

Consequently, to suppress this excitation, the $1\,$Hz spring suspension must be critically damped [as calculated in Fig.~\ref{fig:fig3}(b)]. While the damping of spring-suspended microscopy modules in cryostats has conventionally been achieved using eddy currents~\cite{oh2021cryogen}, the requirement for our SPM setup to operate within a 9\,T magnetic field strictly prohibits the use of the permanent magnets required for this scheme. 

To resolve this conflict, we introduce a novel, fully magnetic-field- and low-temperature-compatible damping method inspired by viscous oil dampers. As shown in Fig.~\ref{fig:fig3}(c), this damping stage consists of an OFHC Cu hook, rigidly attached to the bottom of the suspended AFM module, submerged into an OFHC Cu pot filled with soft copper wool. The pot is rigidly anchored to the MXC attachment via a scaffold structure [Fig.~\ref{fig:fig3}(a)]. Mimicking the role of a highly viscous fluid, the copper wool absorbs and dissipates mechanical energy through deformation, thereby damping the vertical and horizontal deflections of the hook. Furthermore, the hook and pot side walls are outfitted with an interlocking fin structure to constrain the rotational degrees of freedom of the suspended AFM module. By gradually adjusting the amount and packing density of the copper wool, the suspension stage was tuned to critical damping (damping coefficient $c_{\rm AFM}\approx3.3\,\text{kg}/\text{s}$) at room temperature, successfully stabilizing the AFM module against the 1.7\,Hz PT drive. (Note that the damping efficiency may be slightly reduced at cryogenic temperatures owing to the increased stiffness of the copper wool).

\section{\label{sec:results}Results}

\subsection{\label{sec:MIM}Picometer Stability in AFM and sMIM Operation}

We validate the performance of our integrated design by implementing tuning-fork-based sMIM~\cite{cui2016quartz} on the cold-insertable probe, operating at the base temperature of our dilution refrigerator ($T=8\,$mK) with both PT coolers running continuously. For these tests, we examine a test sample that consists of an array of square-shaped Au pads ($4\times4\,\text{µm}^2$, $10\,$nm thick) deposited on a Si(111) substrate.

\begin{figure*}[htbp]
    \centering
    \includegraphics[width=17cm]{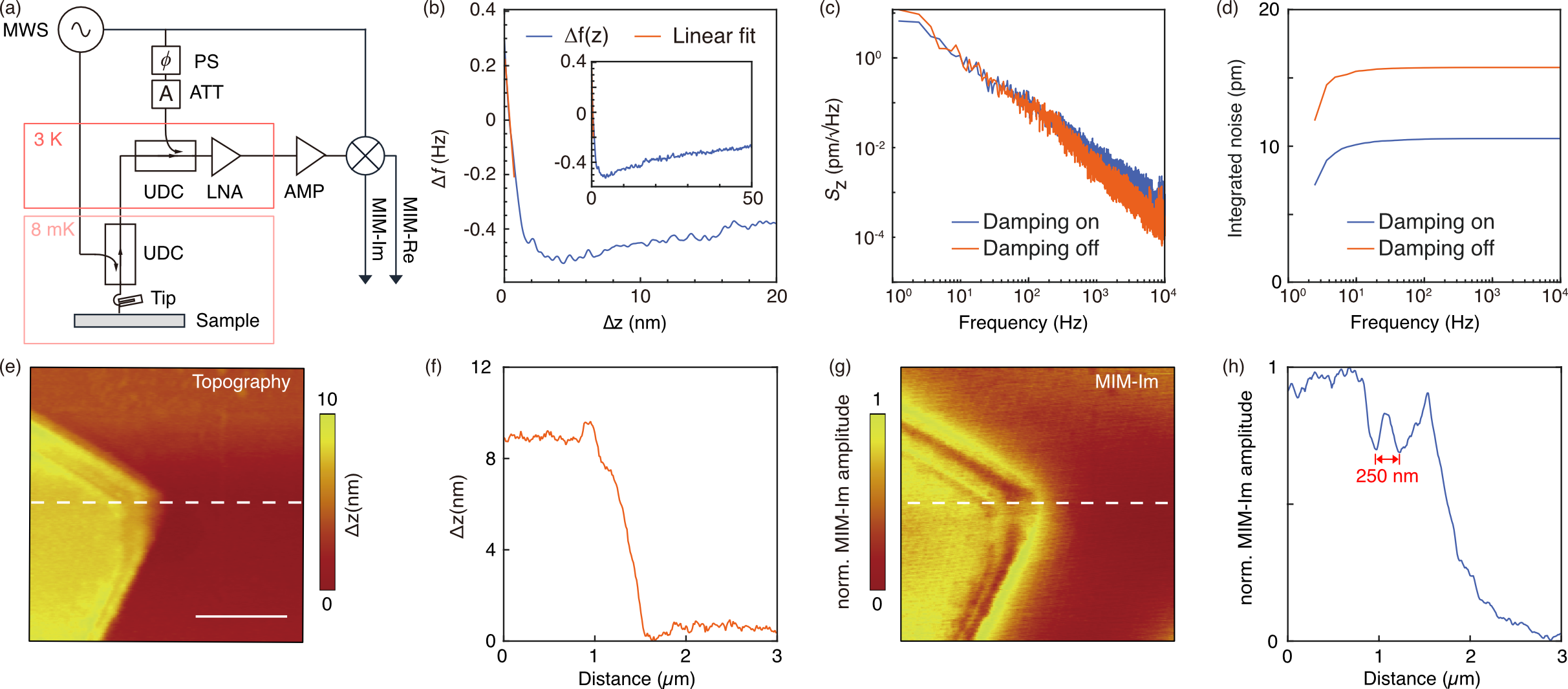}
    \caption{\textbf{Microwave impedance microscopy circuit configuration, noise characterization, and representative imaging results.} 
    \textbf{(a)} Schematic of the sMIM circuit. The output of the microwave source (MWS)~\cite{MWS} is routed via a three-way power splitter to hand-formable microwave cables leading to the mixing chamber, and then via flexible cables~\cite{MWcable} to the AFM head. An impedance matching network (a $0.1\,$pF capacitor to the tip wire) couples the signal to the AFM tip-sample junction~\cite{cui2016quartz}. The reflected measurement signal is separated using a unidirectional coupler (UDC)~\cite{Dcoup} at the mixing chamber. A cancellation line---adjusted via a phase shifter (PS)~\cite{PS} and tunable attenuator (ATT)~\cite{ATT}---suppresses the common mode signal at a second UDC. The signal is amplified by a 3\,K HEMT low-noise amplifier~\cite{LNA} ($40\,$dB) and a room-temperature amplifier~\cite{Amp} ($22\,$dB), then demodulated by an IQ-mixer~\cite{Mixer}. 
    \textbf{(b)} AFM frequency-shift--tip-distance ($\Delta f$--$z$) curve. The inset shows the measurement over the full tip--sample distance range. The red line represents a linear fit to the local response near $\Delta f = 0\,\mathrm{Hz}$. \textbf{(c)} Measured noise spectral density $S_{\rm z}(f)$ of the readout signal as a function of frequency $f$. \textbf{(d)} Frequency-integrated noise amplitude as calculated from $S_{\rm z}(f)$ shown in panel~(c). \textbf{(e)} Representative AFM topography recorded on the test sample. \textbf{(f)}  Line profile extracted from the AFM topography along the dashed line in panel~(e). \textbf{(g)} Representative MIM-Im image acquired over the same field of view as panel~(e). \textbf{(h)} Line profile extracted from the MIM-Im image along the dashed line in panel~(g).}
    \label{fig:fig4}
\end{figure*}

{\em AFM and sMIM Operation:} The coarse and scan motions, as well as tip-sample distance control via frequency-modulation feedback of the AFM, are driven by commercial SPM control electronics~\cite{RHK}. To simultaneously record spatially resolved electric material properties, we implemented a microwave reflectometry circuit [see simplified schematic in Fig.~\ref{fig:fig4}(a)]. The imaginary part of the demodulated microwave signal (MIM-Im) provides a high-resolution measure of the local electric conductivity~\cite{barber2022microwave}. An impedance matching network is integrated directly into the AFM module [(6) in Fig.~\ref{fig:fig2}(b)] to facilitate efficient coupling to the tip-sample junction. The detailed routing, filtering, and amplification scheme of the microwave lines is described in the caption of Fig.~\ref{fig:fig4}.

{\em Characterization of Tip-Sample Vibrations:} To quantify the mechanical stability of the setup, we first calibrate the frequency shift $\Delta f$ of the tuning fork against the tip-sample distance $\Delta z$. The resulting force-distance curve [Fig.~\ref{fig:fig4}(b)] exhibits a steep repulsive transition at $\Delta z<1\,$nm. A linear least-square fit to this region yields a conversion factor of $\alpha=(-1.09\pm0.16)\,\text{Hz}/\text{nm}$, allowing us to calibrate picometer-scale oscillation amplitudes. Next, with the AFM tip stabilized near the sample surface ($\Delta f \approx 0\,$Hz), we open the feedback loop and record the noise spectral density of the read-out signal. Converting this using $\alpha$ yields the spectral noise density of the relative tip-sample distance, $S_{\rm Z}(f)$, shown in Fig.~\ref{fig:fig4}(c). 

Remarkably, $S_{\rm Z}(f)$ is completely free of resonance peaks and exhibits a smooth reduction toward higher frequencies. This featureless spectrum indicates the effectiveness of our approach to spectrally separate and mechanically decouple the structural eigenmodes of the stiff AFM module ($f_{\rm Scanner}$ and $f_{\rm AFM}$) from the noise spectrum of the PT [see Fig.~\ref{fig:fig1}(b)] using a critically damped spring isolation. Consequently, we achieve a spectrally integrated [$10^0-10^4\,$Hz] relative tip-sample vibration of $\Delta z \approx10\,$pm [Fig.~\ref{fig:fig4}(d)]. This contrasts sharply with previous sGM~\cite{oh2021cryogen} and sMIM/sSET~\cite{barber2024characterization} setups in dry refrigerators, where under-damped isolation stages or nanopositioning stages with low-frequency resonance modes resulted in severe resonance peaks of $10^0-10^1\,\text{nm}/\sqrt{\text{Hz}}$ and integrated vibrations between $2$ and $30\,$nm.

Finally, we demonstrate the sMIM and AFM functionality of our setup. In Figs.~\ref{fig:fig4},~(e) and (f), we show an AFM topography and a line cut [white dashed line in panel (e)], respectively, recorded near a gold pad of the test sample. The high stability of our AFM setup results in essentially noise-free data, resolving even the finest sub-nanometer topographic features of the gold pad. In Figs.~\ref{fig:fig4},~(g) and (h), we present a two-dimensional map and line cut [white dashed line in panel (g)] of the imaginary part (MIM-Im) of the detected microwave signal, respectively, which are recorded in the same region as the topographic AFM data. Note that the high stability of the AFM setup facilitates the sMIM imaging of fine conductive features as narrow as 250\,nm.

\subsection{\label{sec:Etemp}Realizing Ultra-Low Electron Temperatures in a Spring-Suspended AFM}

\subsubsection{Thermalization and Wiring Architecture}

The critical trade-off in cryogenic scanning probe microscopy is that the mechanical suspensions required for vibration isolation inherently sever the direct thermal links necessary to cool the sample. Achieving ultra-low electron temperatures ($T_{\rm e} < 100\,$mK) at the AFM sample stage requires overcoming this thermal resistance without stiffening the mechanical compliance of the springs or the Pan-walker. To realize this, we implemented a comprehensive thermalization, wiring, and radio-frequency (RF) filtering architecture on the cold-insertable probe [schematically overviewed in Fig.~\ref{fig:fig5}(a)].

{\em Thermal Strapping:} To maximize thermal conductance, all structural components of the AFM module and MXC attachment are machined from OFHC Cu ($>100\,\text{W}/\text{m}\cdot\text{K}$ at $4\,$K~\cite{valois2024characterization}), mirror-polished via electrochemical etching, and plated with $1\,\text{µm}$ of gold to minimize thermal boundary resistance across joints~\cite{gmelin1999thermal}.

Bridging the moving components requires highly flexible thermal strapping. To thermalize the walker assembly [Fig.~\ref{fig:fig2}(d)] to the microscopy body without impeding its millimeter-scale travel, we developed specialized straps [Fig.~\ref{fig:fig5}(b)] by laser-cutting a longitudinal filament structure into a $250\,\text{µm}$ thick OFHC Cu foil, followed by gold plating. To thermally bypass the $1\,$Hz spring suspension, we developed flexible thermalization braids [Fig.~\ref{fig:fig5}(c)]. These consist of OFHC Cu wire braids ($5\times0.5\,\text{mm}^2$ cross-section, $200\,$mm length) electron-beam-welded~\cite{gmelin1999thermal} into solid, polished, and gold-plated OFHC Cu end pieces, ensuring a robust thermal link directly from the MXC to the AFM module [(1) in Fig.~\ref{fig:fig5}(a)].

{\em Wiring and RF Filtering:} To prevent parasitic heat leaks and electronic heating of the sample, all electrical connections from room temperature (RT) to the MXC utilize low-thermal-conductivity wiring ($12\times$ CuNi twisted pairs for DC/control lines, graphitic coaxial cables for RF signals), anchored at each temperature stage with copper heat sinks. Below the MXC plate, we transition to $100\,\text{µm}$-thin Cu wire~\cite{Wire}, which are further thermalized at the MXC attachment and AFM module body using bobbins. Finally, to eliminate high-frequency electronic noise, all control and signal lines (except the tuning fork) are equipped with RT low-pass $\pi$-filters~\cite{RTfilter}, while the five sample signal lines are additionally outfitted with thermally anchored silver-powder filters~\cite{MXCfilter} at the MXC level.

\subsubsection{Measurement of the Sample Electron Temperature}

\begin{figure*}[htbp]
    \centering
    \includegraphics[width=17cm]{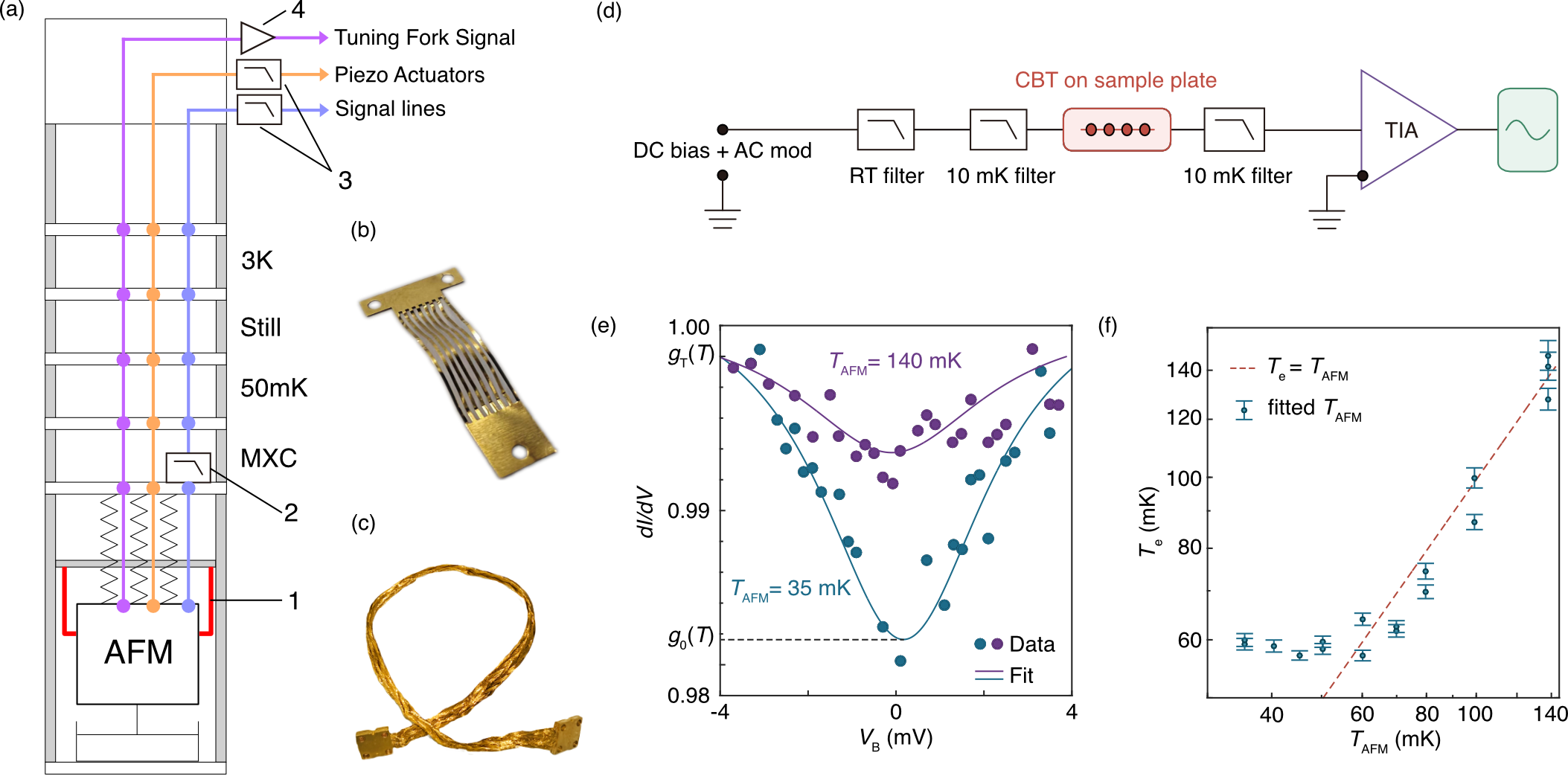}
    \caption{\textbf{Thermalization and filtering architecture of the AFM module and electron-temperature characterization.} \textbf{(a)} Schematic of the thermal links, electrical wiring, and filtering scheme realized on the cold-insertable probe of the dilution refrigerator. The purple, orange, and blue lines denote the AFM signal, piezo-actuator drive lines, and sample signal lines, respectively. The labeled components are the direct thermal link between the mixing-chamber (MXC) stage and the AFM module (1), filters at the 10-mK stage~\cite{MXCfilter} (2), room-temperature filters~\cite{RTfilter} (3), and the AFM signal amplifier~\cite{Femto} (4). \textbf{(b,c)} Home-designed, gold-plated OFHC Cu thermal links: a laser-cut flexible Cu foil thermally anchoring the piezoelectric walker assembly and an electron-beam-welded Cu braid providing a direct thermal link between the MXC attachment and the AFM module. \textbf{(d)} Measurement circuit schematics used for Coulomb-blockade thermometry of the electron temperature $T_{\rm e}$. \textbf{(e)} Representative normalized differential-conductance $dI/dV$ spectra shown as a function of bias voltage at \(T_{\mathrm{AFM}}=35\) and \(140~\mathrm{mK}\). Symbols denote the measured data, and solid curves represent the corresponding fits. \textbf{(f)} Extracted electron temperature \(T_{\rm e}\) as a function of the AFM-module temperature \(T_{\mathrm{AFM}}\). The dashed line indicates \(T_{\rm e}=T_{\mathrm{AFM}}\).}
    \label{fig:fig5}
\end{figure*}

Because the sMIM measurement signal is not directly sensitive to the true electron temperature $T_{\rm e}$, we quantified the cooling efficiency of our setup using a primary Coulomb blockade thermometer (CBT)~\cite{casparis2012metallic} [circuit schematic in Fig.~\ref{fig:fig5}(d)]. 

The CBT sensor was adhesively attached to an OFHC Cu sample plate, inserted into the AFM, and cooled to the dilution refrigerator's base temperature. Using a RuO$_2$ thermometer mounted on the top plate of the AFM module, we recorded a lattice temperature of $T_{\rm AFM}\approx35\,$mK (with the MXC held at $\approx10\,$mK). The bias voltage ($V_{\rm B}$) dependent differential tunnel conductance $dI/dV$ of the CBT was measured using standard lock-in techniques~\cite{LI} with a $4\,\text{µV}$ AC modulation and a transimpedance amplifier~\cite{TIA}.

We recorded $dI/dV$ spectra while incrementally increasing the lattice temperature $T_{\rm AFM}$ from $35\,$mK to $140\,$mK using an MXC heater. Representative spectra [Fig.~\ref{fig:fig5}(e)] exhibit the characteristic zero-bias conductance dip associated with the Coulomb blockade effect. To extract $T_{\rm e}$, we fitted the spectra using a Lorentzian line shape to determine the zero-bias conductance $g_0$ and the normal state conductance $g_T$ (at $V_{\rm B}=-4\,$mV). The temperature $T_{\rm e}$ was then calculated by fitting the ratio $g_0/g_T$ to the standard high-temperature expansion at each $T_{\rm AFM}$ value~\cite{casparis2012metallic}:
\begin{equation*}
    1-\frac{g_0}{g_T}=\frac{u}{6}-\frac{u^2}{60}+\frac{u^3}{630},
\end{equation*}
where $u=2E_{\rm C}/(k_{\rm B} T_{\rm e})$ and $E_{\rm C}$ is the charging energy. 

The extracted electron temperature $T_{\rm e}$ is plotted as a function of the AFM lattice temperature $T_{\rm AFM}$ in Fig.~\ref{fig:fig5}(f). Down to $60\,$mK, the electron temperature perfectly tracks the lattice temperature ($T_{\rm e}\approx T_{\rm AFM}$). Below this point, $T_{\rm e}$ saturates, demonstrating that the sample electrons in our AFM setup reach an effective minimum temperature of $T_{\rm e} \leq 60\,$mK. Crucially, this result confirms that our thermal strapping and filtering architecture successfully cools the local sample environment to the ultra-low temperature regime despite the use of piezoelectric positioning stages and vibration isolation schemes.

\section{\label{sec:DisConc}Discussion and Conclusion}

The experimental characterization of our instrument demonstrates that our two-pronged design strategy---spectrally separating the structural resonances of a stiff AFM and mechanically isolating it via a critically damped internal spring suspension---yields a remarkably small relative tip-sample vibration of $\Delta z \approx 10\,$pm at the base temperature of a dry dilution refrigerator. 

This approach represents a major advancement over existing low-temperature SPM architectures. Historically, achieving sub-Angstrom stability in dry systems required substantial and permanent modifications to the host cryostat, such as physically decoupling the PT coolers and introducing massive isolation stages~\cite{pelliccione2013design, den2014atomic}. Even compared to these heavily modified systems (which achieved vibration amplitudes of roughly $60\,$pm), our design facilitates a six-fold stability improvement while being entirely housed on a standard, motorized cold-insertable probe. Furthermore, in direct comparison to recent AFM-based setups that utilize similar fast-loading probe mechanisms~\cite{cao2023millikelvin, Barber2024}, our rigorous application of mechanical and thermal design principles results in a greater than 100-fold improvement in mechanical stability [see Fig.~\ref{fig:fig1}(c)].

In summary, we have resolved the fundamental trade-off between mechanical vibration isolation and effective thermalization for scanning probe microscopy in cryogen-free environments. By coupling a mechanically stiff AFM module ($f_0>2\,$kHz) with an internal, critically damped spring-isolation stage ($f_{\rm Spring}\approx1\,$Hz), we suppressed the spectrally integrated tip-sample displacement to $\approx10\,$pm ($S_{\rm AFM}(f)<10^{-11}\,\text{m}/\sqrt{\text{Hz}}$). Simultaneously, our optimized thermal interfaces, custom copper strapping, and comprehensive RF filtering enabled the local electron temperature to reach $T_{\rm e}\leq60\,$mK. 

These results establish that sub-Angstrom stability---sufficient for resolving individual atoms and molecules~\cite{giessibl2000subatomic, gross2009chemical}---and ultra-low electron temperatures can coexist despite the intense mechanical noise of PT coolers. Moving forward, this modular architecture provides a highly accessible framework for integrating other scanning probe techniques, such as sSQUID and sSET into a dry dilution refrigerator for picometer-stable operation. By eliminating the need for dedicated, permanently modified cryostats, our design accelerates and broadens the nanoscale exploration of fragile quantum phases in fast-turnaround, dry experimental setups.

\begin{acknowledgments}
We are grateful for the support of Sasha Usenko and David Qiu from Leiden Cryogenics for the design and installation of the dry dilution refrigerator, FOCUS GmbH for the joint development of electron-beam welded thermal straps, and Dominik Zumbuehl and Basel Precision Instruments for the provision and advice on the operation of the Coulomb blockade thermometer. This work was supported by the Croucher Foundation through Grant No.\,CIA22SC02.
\end{acknowledgments}

\section*{Data Availability Statement}

The data that support the findings of this study are available from the corresponding author upon reasonable request.

%\appendix

%\section{Appendixes}

%\setcounter{figure}{0}
%\renewcommand{\thefigure}{A\arabic{figure}}

\bibliography{bibliography}% Produces the bibliography via BibTeX.

@PREAMBLE{
 "\providecommand{\noopsort}[1]{}" 
 # "\providecommand{\singleletter}[1]{#1}%" 
}

@article{Telford2023,
  author  = {Telford, E. J. and Ben-Shalom, M. and others},
  title   = {MilliKelvin microwave impedance microscopy in a dry dilution refrigerator},
  journal = {Review of Scientific Instruments},
  volume  = {94},
  number  = {9},
  pages   = {093705},
  year    = {2023},
  doi     = {10.1063/5.0135805}
}

@article{Pelliccione2013,
  author  = {Pelliccione, M. and Sciambi, A. and Bartel, J. and Keller, A. J. and Goldhaber-Gordon, D.},
  title   = {Design of a scanning gate microscope in a cryogen-free dilution refrigerator},
  journal = {Review of Scientific Instruments},
  volume  = {84},
  number  = {3},
  pages   = {033703},
  year    = {2013},
  doi     = {10.1063/1.4794767}
}

@article{Ge2025,
  author  = {Ge, Z. and others},
  title   = {A milli-Kelvin atomic force microscope made of glass},
  journal = {arXiv preprint arXiv:2502.19845},
  year    = {2025},
  url     = {https://arxiv.org/abs/2502.19845}
}

@article{assig201310,
  title={A 10 mK scanning tunneling microscope operating in ultra high vacuum and high magnetic fields},
  author={Assig, Maximilian and Etzkorn, Markus and Enders, Axel and Stiepany, Wolfgang and Ast, Christian R and Kern, Klaus},
  journal={Review of Scientific Instruments},
  volume={84},
  number={3},
  year={2013},
  publisher={AIP Publishing}
}

@article{ast2008design,
  title={Design criteria for scanning tunneling microscopes to reduce the response to external mechanical disturbances},
  author={Ast, Christian R and Assig, Maximilian and Ast, Alexandra and Kern, Klaus},
  journal={Review of Scientific Instruments},
  volume={79},
  number={9},
  year={2008},
  publisher={AIP Publishing}
}

@article{oh2021cryogen,
  title={Cryogen-free scanning gate microscope for the characterization of Si/Si0. 7Ge0. 3 quantum devices at milli-Kelvin temperatures},
  author={Oh, Seong Woo and Denisov, Artem O and Chen, Pengcheng and Petta, Jason R},
  journal={AIP Advances},
  volume={11},
  number={12},
  year={2021},
  publisher={AIP Publishing}
}

@article{denHaan2014,
  author  = {den Haan, A. M. J. and Guallart-Naval, T. and others},
  title   = {Atomic resolution scanning tunneling microscopy in a cryogen free dilution refrigerator at 15 mK},
  journal = {Review of Scientific Instruments},
  volume  = {85},
  number  = {3},
  pages   = {035112},
  year    = {2014},
  doi     = {10.1063/1.4868684}
}

@article{keimer2017physics,
  title={The physics of quantum materials},
  author={Keimer, Bernhard and Moore, Joel E},
  journal={Nature Physics},
  volume={13},
  number={11},
  pages={1045--1055},
  year={2017},
  publisher={Nature Publishing Group UK London}
}

@article{persky2022studying,
  title={Studying quantum materials with scanning SQUID microscopy},
  author={Persky, Eylon and Sochnikov, Ilya and Kalisky, Beena},
  journal={Annual Review of Condensed Matter Physics},
  volume={13},
  number={1},
  pages={385--405},
  year={2022},
  publisher={Annual Reviews}
}

@article{barber2022microwave,
  title={Microwave impedance microscopy and its application to quantum materials},
  author={Barber, Mark E and Ma, Eric Yue and Shen, Zhi-Xun},
  journal={Nature Reviews Physics},
  volume={4},
  number={1},
  pages={61--74},
  year={2022},
  publisher={Nature Publishing Group UK London}
}

@article{Lee2022,
  author  = {Lee, J. H. and others},
  title   = {Development of a near-5-Kelvin, cryogen-free, pulse-tube refrigerator-based scanning probe microscope},
  journal = {Review of Scientific Instruments},
  volume  = {93},
  number  = {8},
  pages   = {083702},
  year    = {2022},
  doi     = {10.1063/5.0097723}
}

@article{geaney2019near,
  title={Near-field scanning microwave microscopy in the single photon regime},
  author={Geaney, S and Cox, D and H{\"o}nigl-Decrinis, T and Shaikhaidarov, R and Kubatkin, SE and Lindstr{\"o}m, T and Danilov, AV and de Graaf, SE},
  journal={Scientific reports},
  volume={9},
  number={1},
  pages={12539},
  year={2019},
  publisher={Nature Publishing Group UK London}
}

@article{Barber2024,
  author  = {Barber, M. E. and others},
  title   = {Characterization of Two Fast-Turnaround Dry Dilution Refrigerators for Scanning Probe Microscopy},
  journal = {Journal of Low Temperature Physics},
  year    = {2024},
  doi     = {10.1007/s10909-024-03102-1}
}

@article{battisti2018definition,
  title={Definition of design guidelines, construction, and performance of an ultra-stable scanning tunneling microscope for spectroscopic imaging},
  author={Battisti, Irene and Verdoes, Gijsbert and van Oosten, Kees and Bastiaans, Koen M and Allan, Milan P},
  journal={Review of Scientific Instruments},
  volume={89},
  number={12},
  year={2018},
  publisher={AIP Publishing}
}

@article{white2011stiff,
  title={A stiff scanning tunneling microscopy head for measurement at low temperatures and in high magnetic fields},
  author={White, SC and Singh, UR and Wahl, P},
  journal={Review of Scientific Instruments},
  volume={82},
  number={11},
  year={2011},
  publisher={AIP Publishing}
}

@article{pan1999he,
  title={He 3 refrigerator based very low temperature scanning tunneling microscope},
  author={Pan, SH and Hudson, Eric W and Davis, JC},
  journal={Review of scientific instruments},
  volume={70},
  number={2},
  pages={1459--1463},
  year={1999},
  publisher={American Institute of Physics}
}

@misc{PTcooler,
    key = PTcooler,
    note = {RP-182 from Sumitomo Heavy Industries}
}

@misc{Zpiezo,
    key = Zpiezo,
    note = {P-121.01T from PI Ceramics}
}

@misc{XYpiezo,
    key = XYpiezo,
    note = {PAXZ+0049 from PI Ceramics}
}

@misc{Wire,
    key = Wire,
    note = {311-KAP-010-10M from Allectra}
}

@article{casparis2012metallic,
  title={Metallic Coulomb blockade thermometry down to 10 mK and below},
  author={Casparis, L and Meschke, Matthias and Maradan, D and Clark, AC and Scheller, CP and Schwarzw{\"a}lder, KK and Pekola, Jukka P and Zumb{\"u}hl, DM},
  journal={Review of Scientific Instruments},
  volume={83},
  number={8},
  year={2012},
  publisher={AIP Publishing}
}

@misc{MXCfilter,
    key = MXCfilter,
    note = {slimMFT8 from Basel Precision Instruments}
}

@misc{RTfilter,
    key = RTfilter,
    note = {51-726-001 from Spectrum Control}
}

@misc{Ptube,
    key = Ptube,
    note = {Type 2 from EBL Products}
}

@misc{TF,
    key = TF,
    note = {DS26 from Micro Crystal}
}

@misc{VibIso,
    key = VibIso,
    note = {Halcyonics Duo 73}
}

@misc{MWS,
    key = MWS,
    note = {SRS SG386 from Stanford Research Systems}
}

@misc{PS,
    key = PS,
    note = {ZN3PD-622W-S+ from Mini Circuits}
}

@misc{MWcable,
    key = MWcable,
    note = {PE-P047 from Pasternack}
}

@misc{Dcoup,
    key = Dcoup,
    note = {ZUDC10-0283-S+ from Mini Circuits}
}

@misc{ATT,
    key = ATT,
    note = {PE7436 from Pasternack}
}

@misc{LNA,
    key = LNA,
    note = {WHF-C-B-009 CETC16 from SpinQ}
}

@misc{Amp,
    key = Amp,
    note = {ZX60-14LN-S+ from Mini Circuits}
}

@misc{LI,
    key = LI,
    note = {SR830 from Stanford Research Systems}
}

@misc{TIA,
    key = TIA,
    note = {SSP983c from Basel Precision Instruments}
}

@article{gross2009chemical,
  title={The chemical structure of a molecule resolved by atomic force microscopy},
  author={Gross, Leo and Mohn, Fabian and Moll, Nikolaj and Liljeroth, Peter and Meyer, Gerhard},
  journal={Science},
  volume={325},
  number={5944},
  pages={1110--1114},
  year={2009},
  publisher={American Association for the Advancement of Science}
}

@article{giessibl2000subatomic,
  title={Subatomic features on the silicon (111)-(7$\times$ 7) surface observed by atomic force microscopy},
  author={Giessibl, Franz J and Hembacher, Stefan and Bielefeldt, Hartmut and Mannhart, Jochen},
  journal={Science},
  volume={289},
  number={5478},
  pages={422--425},
  year={2000},
  publisher={American Association for the Advancement of Science}
}

@misc{Mixer,
    key = Mixer,
    note = {MMIQ-0218HXPC from Marki Microwave}
}

@misc{RHK,
    key = RHK,
    note = {R10 from RHK}
}

@misc{Femto,
    key = Femto,
    note = {HQA-15M-10T from Femto}
}

@article{jiang2023implementing,
  title={Implementing microwave impedance microscopy in a dilution refrigerator},
  author={Jiang, Zhanzhi and Chong, Su Kong and Zhang, Peng and Deng, Peng and Chu, Shizai and Jahanbani, Shahin and Wang, Kang L and Lai, Keji},
  journal={Review of Scientific Instruments},
  volume={94},
  number={5},
  year={2023},
  publisher={AIP Publishing}
}

@article{gmelin1999thermal,
  title={Thermal boundary resistance of mechanical contacts between solids at sub-ambient temperatures},
  author={Gmelin, E and Asen-Palmer, M and Reuther, M and Villar, R},
  journal={Journal of Physics D: Applied Physics},
  volume={32},
  number={6},
  pages={R19--R43},
  year={1999}
}

@article{pelliccione2013design,
  title={Design of a scanning gate microscope for mesoscopic electron systems in a cryogen-free dilution refrigerator},
  author={Keller, AJ},
  journal={Review of Scientific Instruments},
  volume={84},
  number={3},
  year={2013},
  publisher={AIP Publishing}
}

@article{cao2023millikelvin,
  title={MilliKelvin microwave impedance microscopy in a dry dilution refrigerator},
  author={Cao, Leonard Weihao and Wu, Chen and Bhattacharyya, Rajarshi and Zhang, Ruolun and Allen, Monica T},
  journal={Review of Scientific Instruments},
  volume={94},
  number={9},
  year={2023},
  publisher={AIP Publishing}
}

@article{finkler2012scanning,
  title={Scanning superconducting quantum interference device on a tip for magnetic imaging of nanoscale phenomena},
  author={Finkler, Amit and Vasyukov, Denis and Segev, Yair and Ne'eman, Lior and Lachman, EO and Rappaport, ML and Myasoedov, Yuri and Zeldov, Eli and Huber, ME},
  journal={Review of Scientific Instruments},
  volume={83},
  number={7},
  year={2012},
  publisher={AIP Publishing}
}

@inproceedings{valois2024characterization,
  title={Characterization of the thermal properties of OFHC copper at cryogenic temperature},
  author={Valois, JJ and Nellis, GF and Pfotenhauer, JM},
  booktitle={IOP conference series: materials science and engineering},
  volume={1301},
  number={1},
  pages={012167},
  year={2024},
  organization={IOP Publishing}
}

@article{cui2016quartz,
  title={Quartz tuning fork based microwave impedance microscopy},
  author={Cui, Yong-Tao and Ma, Eric Yue and Shen, Zhi-Xun},
  journal={Review of Scientific Instruments},
  volume={87},
  number={6},
  year={2016},
  publisher={AIP Publishing}
}

@article{giessibl2019qplus,
  title={The qPlus sensor, a powerful core for the atomic force microscope},
  author={Giessibl, Franz J},
  journal={Review of scientific instruments},
  volume={90},
  number={1},
  pages={011101},
  year={2019},
  publisher={AIP Publishing LLC}
}

@article{den2014atomic,
  title={Atomic resolution scanning tunneling microscopy in a cryogen free dilution refrigerator at 15 mK},
  author={Den Haan, AMJ and Wijts, GHCJ and Galli, F and Usenko, O and Van Baarle, GJC and Van Der Zalm, DJ and Oosterkamp, TH},
  journal={Review of Scientific Instruments},
  volume={85},
  number={3},
  year={2014},
  publisher={AIP Publishing}
}

@article{barber2024characterization,
  title={Characterization of two fast-turnaround dry dilution refrigerators for scanning probe microscopy},
  author={Barber, Mark E and Li, Yifan and Gibson, Jared and Yu, Jiachen and Jiang, Zhanzhi and Hu, Yuwen and Ji, Zhurun and Nandi, Nabhanila and Hoke, Jesse C and Bishop-Van Horn, Logan and others},
  journal={Journal of Low Temperature Physics},
  volume={215},
  number={1},
  pages={1--23},
  year={2024},
  publisher={Springer}
}

@article{low2021scanning,
  title={Scanning SQUID microscopy in a cryogen-free dilution refrigerator},
  author={Low, David and Ferguson, GM and Jarjour, Alexander and Schaefer, Brian T and Bachmann, Maja D and Moll, Philip JW and Nowack, Katja C},
  journal={Review of Scientific Instruments},
  volume={92},
  number={8},
  year={2021},
  publisher={AIP Publishing}
}

\end{document}